\documentclass{cimento}

\usepackage[utf8]{inputenc}
\usepackage{amsmath}
\usepackage{amssymb}
\usepackage{graphicx}

\newcommand{\fref}[1]{Fig. \ref{#1}}

\title{Glueballs from Dyson-Schwinger and Bethe-Salpeter equations}
\author{M.~Q.~Huber\from{ins:JLU}\ETC,
C.~S.~Fischer\from{ins:JLU}\from{ins:HFHF}
\atque
H.~Sanchis-Alepuz\from{ins:SAL}
}
\instlist{
\inst{ins:JLU} Institut f\"ur Theoretische Physik, Justus-Liebig--Universit\"at Gie\ss en, 35392 Gie\ss en, Germany
\inst{ins:HFHF} Helmholtz Forschungsakademie Hessen f\"ur FAIR (HFHF), GSI Helmholtzzentrum f\"ur Schwerionenforschung, Campus Gie{\ss}en, 35392 Gie{\ss}en, Germany
\inst{ins:SAL} Silicon Austria Labs GmbH, Campus TU Graz, Sandgasse 34, 8010 Graz, Austria}

\begin{document}

\maketitle

\begin{abstract}
The quenched spectrum of glueballs with positive charge parity is calculated from two-body bound state equations.
As input, a self-contained solution for the primitively divergent correlation functions from Dyson-Schwinger equations is used.
It has only one parameter to be set which is the physical scale.
An important feature of this setup is the consistent construction of the bound state kernels along the same lines as the equations from which the input was obtained.
Keeping only the one-particle exchanges, already good agreement with lattice results is obtained.
For the tensor glueball, we present first results including two-loop contributions, elevating its calculation to the same level of truncation as for the spin zero glueballs for which such calculations have been done previously.
\end{abstract}

\section{Introduction}
\label{sec:introduction}

Glueballs are elusive objects that were suggested as bound states of gluons already in the early days of QCD and searched for since then.
Most recent analyses of BESIII data put their lightest realization, the scalar glueball, in the region around 1700 MeV \cite{Sarantsev:2021ein,Rodas:2021tyb}.
A complication in its identification is that several states with scalar quantum numbers exist above 1 GeV.
This mixing makes analyses of experimental data challenging as well as theoretical calculations.

In the functional approach used here, such a mixing arises naturally by the coupling of the amplitudes of two-gluon and quark-antiquark states, see \cite{Huber:2022dsn} for the corresponding equations.
However, there is no solution for full QCD yet.
For pure QCD, on the other hand, viz., when quarks are made infinitely heavy and thus only purely gluonic bound states remain, the situation is as follows.
From lattice calculations the spectrum of the lightest states is known for quite some time \cite{Morningstar:1999rf}.
More recent calculations improved the errors and added some further states \cite{Athenodorou:2020ani}, see also \cite{Vadacchino:2023vnc} for the current status.
For functional methods it was challenging to reproduce this spectrum, although they are successful in describing, for example, the spectrum of baryons or tetraquarks, see \cite{Eichmann:2016yit,Eichmann:2020oqt} and references therein.
One of the main reasons is the truncations necessary for functional bound state equations.
For the quark sector, a particularly simple truncation, called rainbow-ladder truncation, works in many cases quite well.
It only needs to be complemented by an effective quark-gluon interaction.
This truncation works reliably and produces stable results.
In the gluonic sector, first of all a rainbow-like truncation as for the quark propagator does not exist.
Secondly, attempts to model the interactions did not lead to predictive results \cite{Meyers:2012ka,Sanchis-Alepuz:2015hma,Souza:2019ylx,Kaptari:2020qlt}.
The situation is different when input is used that is calculated from a high-level truncation as done in \cite{Huber:2020ngt,Huber:2021yfy}.

This contribution provides an overview of the functional setup including tests for extensions of it in Sec.~\ref{sec:setup}.
The resulting spectrum is presented in Sec.~\ref{sec:results} where also the importance of higher diagrams in the truncation is discussed.
We close with a summary.

\section{Setup}
\label{sec:setup}

\begin{figure}
\begin{center}
 \includegraphics[height=1.4cm]{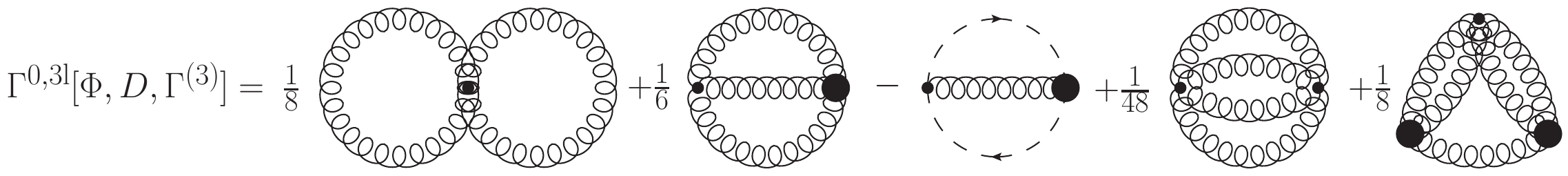}\\

 \includegraphics[height=1.4cm]{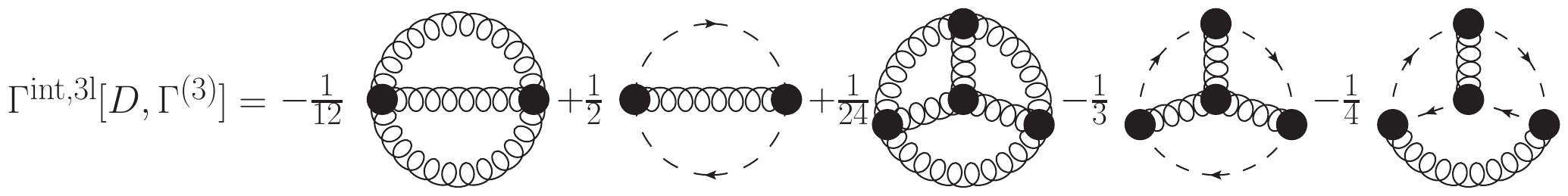}\hspace{1.55cm}
 \end{center}
\caption{$\Gamma^0[\Phi, D, \Gamma^{(3)}]$ and $\Gamma^\text{int}[D, \Gamma^{(3)}]$ of the 3PI effective action truncated to three loops (as indicated by the superscript $\text{3l}$).
Internal propagators are dressed, black disks denote dressed vertices, dots bare vertices, wiggly lines gluons, and dashed lines ghosts. Quark contributions (not shown) have the same structure as the ghost contributions.}
\label{fig:3PI-3L}
\end{figure}

The central quantity in our setup is the 3PI effective action \cite{Berges:2004pu}
\begin{align}
 \Gamma[\Phi, D, \Gamma^{(3)}] &= S[\Phi] + \frac{1}{2}\ln D^{-1}_{ii} + \frac{1}{2} S_{ij}[\Phi] D_{ij} - \Gamma^0[\Phi,D, \Gamma^{(3)}] - \Gamma^\text{int}[D, \Gamma^{(3)}].
\end{align}
$\Phi$ represents the gluon, quark and ghost fields of QCD, $D$ their propagators and $\Gamma^{(3)}$ their three-point functions.
In the case of pure Yang-Mills theory, the quarks contributions are dropped.
The quantities $\Gamma^0[\Phi,D, \Gamma^{(3)}]$ and $\Gamma^\text{int}[D, \Gamma^{(3)}]$ truncated at three loops are depicted in \fref{fig:3PI-3L}.
This is the underlying truncation and can be systematically expanded by higher loop orders.
Note that three loops are necessary for a consistent description of three-point functions \cite{Berges:2004pu}.

\begin{figure}[tb]
	\includegraphics[width=0.6\textwidth]{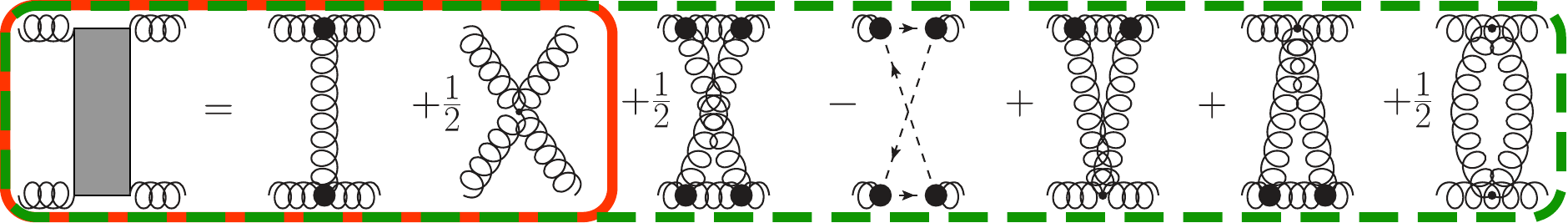}\\ 
	\vskip4mm
	\includegraphics[height=1.1cm]{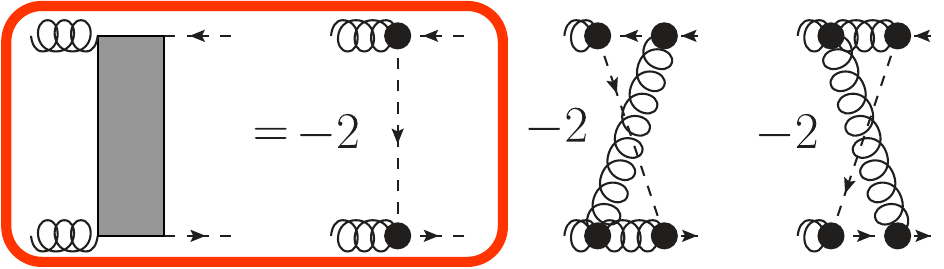}\hfill
	\includegraphics[height=1.1cm]{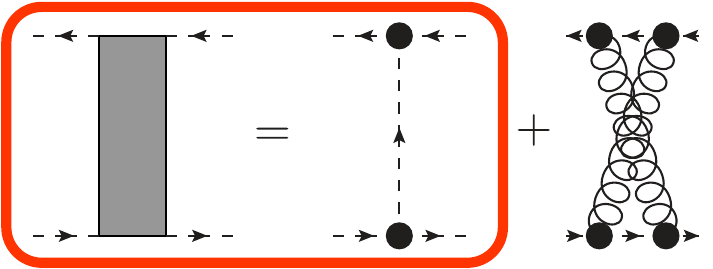}\hfill
	\includegraphics[height=1.1cm]{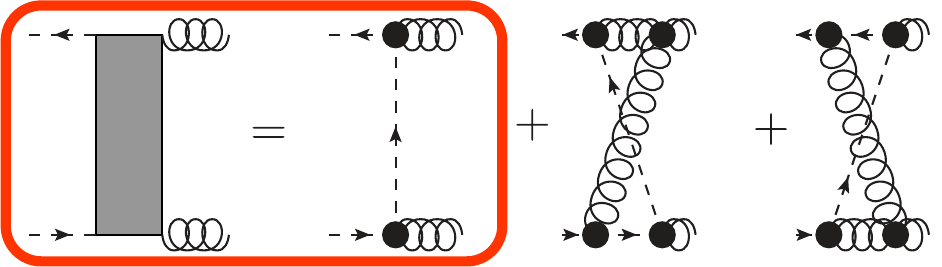}
	\caption{
		Interaction kernels from the three-loop 3PI effective action.
		All propagators are dressed; black disks represent dressed vertices.
		See text for red and green rectangles.
		\label{fig:kernels}
	}
\end{figure}

From the 3PI effective action equations of motion for propagators and three-point functions and expressions for the scattering kernels of the bound state equations can be derived by cutting lines \cite{Huber:2020ngt,Fukuda:1987su,Sanchis-Alepuz:2015tha}.
The latter are shown in \fref{fig:kernels}.
They are to be inserted in the corresponding bound state equations, see \cite{Huber:2020ngt} for details.

The input was calculated in \cite{Huber:2020keu} from the equations of motion of the same effective action and we refer there for details.
We want to emphasize that the input is the result of a long line of calculations of elementary correlation functions, see \cite{Huber:2018ned} and references therein.
The setup finally used contains all correlation functions as dynamic quantities and no model parameters.
The only parameter is the scale.
The following tests of the underlying truncation were performed:
1. The importance of additional dressings for the three-gluon vertex was found to be very small \cite{Eichmann:2014xya}.
2. The impact of the four-gluon vertex was tested in three dimensions and found to be small \cite{Huber:2018ned,Huber:2016tvc}.
3. The impact of other four-point functions was explicitly tested as well and found to be small \cite{Huber:2018ned,Huber:2017txg}.
In addition, the results agree very well with corresponding ones from the functional renormalization group \cite{Cyrol:2016tym}.
All these tests indicate that the input is obtained from a stable truncation and quantitatively reliable.

For solving the bound state equations, the input is needed at time-like momenta.
Instead of an analytic continuation of the corresponding dressing functions, we calculate the eigenvalue of the equations for space-like total momentum $P$.
The obtained eigenvalues are then extrapolated using the Schlessinger continued fraction method \cite{Schlessinger:1968spm}.
The reliability of such an extrapolation was tested for a test system solvable at time-like momenta \cite{Huber:2020ngt}.
The physical solution is where the eigenvalue is one.

\section{Results}
\label{sec:results}

\begin{figure}[tb]
	\begin{center}
	\includegraphics[width=0.47\textwidth]{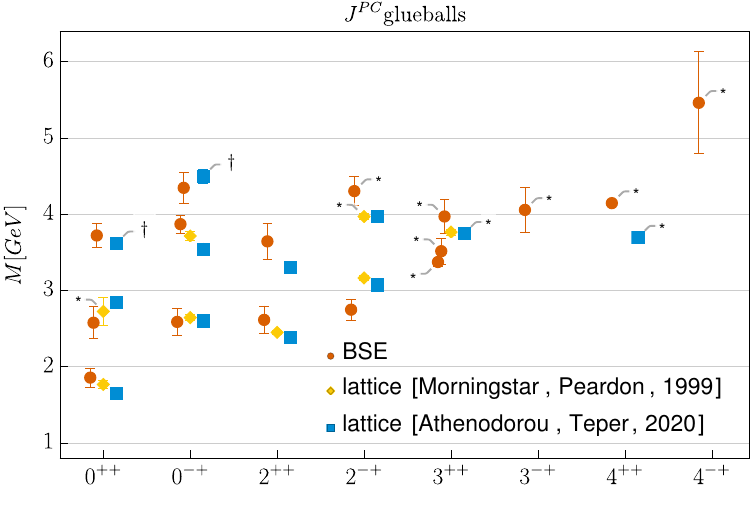}\hfill
	\includegraphics[width=0.47\textwidth]{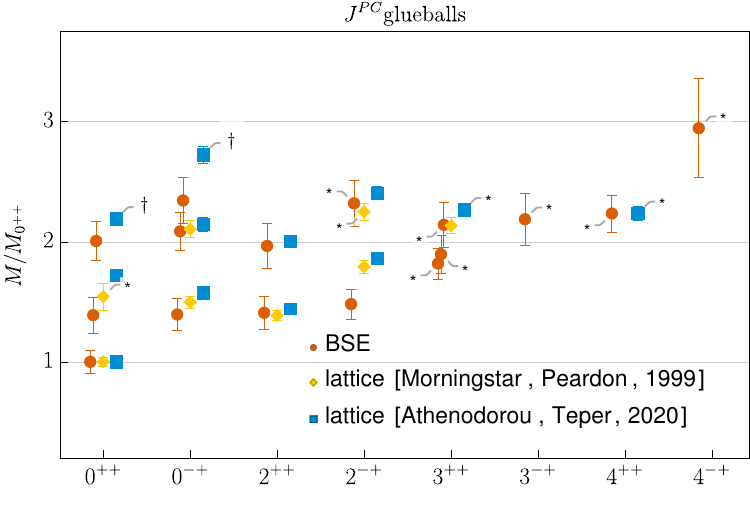}
	\end{center}
	\caption{
		Results for glueball ground states and excited states for the indicated quantum numbers from lattice simulations \cite{Morningstar:1999rf,Athenodorou:2020ani} and functional equations \cite{Huber:2021yfy}.
		In the left plot, we display the glueball masses on an absolute scale set by $r_0=1/418(5)\,\text{MeV})$.
	    In the right plot, we display the spectrum relative to the ground state.
	    Masses with $^\dagger$ are conjectured to be the second excited states.
		Masses with $^*$ come with some uncertainty in their identification in the lattice case or in the trustworthiness of the extrapolated value in the BSE case.
		}
	\label{fig:spectrum}
\end{figure}

Using the kernels depicted in the red rectangles of \fref{fig:kernels}, the spectrum shown in \fref{fig:spectrum} was obtained.
No result could be obtained for $J=1$, supporting the expectation that such states would correspond to three-gluon amplitudes.
However, we stress that the Landau-Yang theorem usually credited for this does not apply here as the gluons are not on-shell.

As a remarkable finding we obtained for the $J=0$ glueballs the same masses independent of the chosen input \cite{Huber:2020ngt}.
Such different solutions arise as a one-parameter family of solutions \cite{Boucaud:2008ji,Aguilar:2008xm,Fischer:2008uz,Alkofer:2008jy}.
Finding the same spectrum from markedly different input supports the conjecture that these solutions are related to the Gribov problem and incomplete gauge fixing \cite{Fischer:2008uz,Maas:2009se}.

Finally, we explored the effect of higher-loop diagrams in the kernels.
To this end, we included the diagrams in the green dashed rectangle in \fref{fig:kernels}.
They belong to the gluon-gluon scattering kernel.
The corresponding one-particle exchange provides the leading contribution in the original truncation.
Due to the more demanding numerical calculation, we solved the equation with smaller precision.
As a consequence, we could only extract the ground state.
For quantitative comparisons with the original truncation, we repeated those calculations at the same level of precision.
For the scalar \cite{Huber:2022rhh}, pseudoscalar \cite{Huber:2021zqk} and tensor glueballs the results are hardly affected.
The largest effect, of the order of $2\%$, was found for the scalar glueball.
However, this is well below the error from other sources like the extrapolation.
For the tensor glueball this extension of the truncation is presented here for the first time and found to be at the sub-per mille level for the eigenvalues.
The resulting mass value is consequently also nearly identical.
By repeating the calculation with only subsets of two-loop diagrams we confirmed that their impact is individually small.
This is in contrast to the calculation of gluonic vertices where considerable cancelations between the diagrams take place \cite{Huber:2016tvc,Cyrol:2014kca}.

\section{Summary}
\label{sec:summary}

We presented results for the glueball spectrum of pure QCD from functional equations.
The employed setup has as only parameter the physical scale, as all contained correlations functions are calculated from their equations of motion.
The obtained masses, which are in good agreement with corresponding results from lattice calculations, can thus be considered as obtained form first principles.
The scattering kernels of the bound states were originally restricted to one-particle exchanges.
This approximation was partially lifted for the three lightest states for which the two-loop diagrams in the gluon scattering kernel were included.
The resulting differences are hardly visible for the pseudoscalar and the tensor (below per mille in the eigenvalues) and small for the scalar (below 2\%).
Consequently, errors from the extrapolation are much larger than errors from neglecting these diagrams.
Besides this extension of the bound state calculation, also several tests for the employed input were discussed which all support its quantitative rigor.

\acknowledgments

This work was supported by the DFG (German Research Foundation) grant FI 970/11-1, by the BMBF under contract No. 05P21RGFP3, and by Silicon Austria Labs (SAL), owned by the Republic of Austria, the Styrian Business Promotion Agency (SFG), the federal state of Carinthia, the Upper Austrian Research (UAR), and the Austrian Association for the Elec­tric and Elec­tronics Industry (FEEI).

\bibliographystyle{utphys_mod}
\bibliography{literature_glueballs_HADRON2023}

\end{document}